\documentclass[conference, a4paper]{IEEEtran}
\IEEEoverridecommandlockouts
\usepackage{cite}
\usepackage{amsmath,amssymb,amsfonts}
\usepackage{algorithmic}
\usepackage{graphicx}
\usepackage{textcomp}
\usepackage{xcolor}
\def\BibTeX{{\rm B\kern-.05em{\sc i\kern-.025em b}\kern-.08em
    T\kern-.1667em\lower.7ex\hbox{E}\kern-.125emX}}

\usepackage{import, color} %
\usepackage{siunitx}
\usepackage{multirow}
\usepackage{tabularx}

\definecolor{LMSred}{rgb}{0.80,0.20,0.20}

\IEEEpubid{\begin{minipage}[t]{\textwidth}\ \\[10pt] 978-1-6654-3288-7/21/\$31.00~\copyright2021 IEEE\end{minipage}}

\usepackage[firstpage=true]{background}
\usepackage{hyperref}

\SetBgContents{\parbox{\textwidth}{\footnotesize © 2021 IEEE. Personal use of this material is permitted. Permission from IEEE must be
		obtained for all other uses, in any current or future media, including
		reprinting/republishing this material for advertising or promotional purposes, creating new
		collective works, for resale or redistribution to servers or lists, or reuse of any copyrighted
		component of this work in other works. DOI: \href{https://doi.org/10.1109/MMSP53017.2021.9733541}{10.1109/MMSP53017.2021.9733541.}}}
\SetBgScale{1}
\SetBgAngle{0}
\SetBgPosition{current page.south}
\SetBgVshift{1cm}
\SetBgColor{black}
\SetBgOpacity{1}

\begin{document}

\title{A Novel End-To-End Network for Reconstruction of Non-Regularly Sampled Image Data Using Locally Fully Connected Layers}

\author{\IEEEauthorblockN{Simon Grosche, Fabian Brand, and André Kaup}
	\IEEEauthorblockA{\textit{Multimedia Communications and Signal Processing}\\
		\textit{Friedrich-Alexander University Erlangen-Nürnberg (FAU)}\\
		Cauerstr. 7, 91058 Erlangen, Germany\\
		\{simon.grosche, fabian.brand, andre.kaup\}@fau.de}
}
\maketitle

\begin{abstract}
Quarter sampling and three-quarter sampling are novel sensor concepts that enable the acquisition of higher resolution images without increasing the number of pixels. This is achieved by non-regularly covering parts of each pixel of a low-resolution sensor such that only one quadrant or three quadrants of the sensor area of each pixel is sensitive to light. 
Combining a properly designed mask and a high-quality reconstruction algorithm, a higher image quality can be achieved than using a low-resolution sensor and  subsequent upsampling.
For the latter case, the image quality  can be  further enhanced using super resolution algorithms such as the very deep super resolution network (VDSR).
In this paper,  we propose a novel end-to-end neural network to reconstruct high resolution images from non-regularly sampled sensor data. The network is a concatenation  of a locally fully connected reconstruction network (LFCR) and a standard VDSR network.
Altogether, using a three-quarter sampling sensor with our novel neural network layout, the image quality in terms of PSNR for the Urban100 dataset can be increased by 2.96\,dB compared to the state-of-the-art approach. Compared to a low-resolution sensor with VDSR, a gain of 1.11\,dB is achieved.
\end{abstract}


\section{Introduction}

\label{sec:intro}
\begin{figure}[t!]
	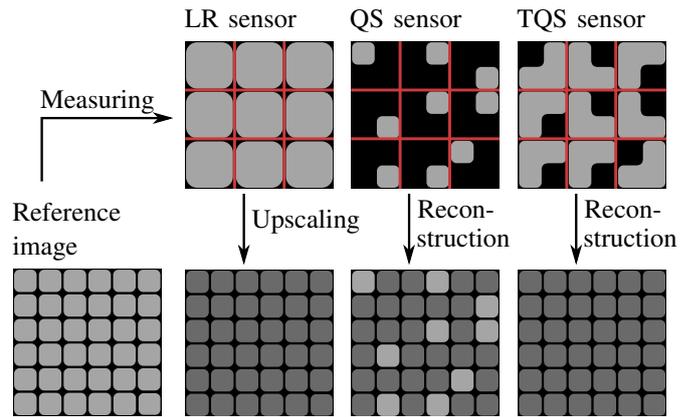
	\vspace*{-2mm}
	\caption{Illustration of an image acquisition using a low-resolution (LR) sensor, a quarter sampling (QS) sensor \cite{Schoberl2011}, and a three-quarter sampling (TQS) sensor \cite{Seiler2018}. Light gray pixels indicate measured pixels whereas dark gray pixels indicate upscaled/reconstructed pixels.}
	\label{fig:LR_vs_quarter}
	\vspace*{-4mm}
\end{figure}

Using quarter sampling  \cite{Schoberl2011}, the spatial resolution of an imaging sensor can be increased. This is achieved by physically covering three quarters of each pixel of a low-resolution sensor as it is illustrated in Figure\,\ref{fig:LR_vs_quarter}. Effectively, this leads to a non-regular sampling of the image with respect to a higher resolution grid with twice the resolution in both spatial dimensions.
Since the sampling is non-regular, it leads to reduced visible aliasing artifacts conventionally occurring for regular sampling \cite{Dippe1985, Hennenfent2007, Maeda2009}.
The missing pixels need to be reconstructed from the sampled data.
For a proper reconstruction, high-quality reconstruction algorithms such as the frequency selective reconstruction (FSR) \cite{Seiler2015} need to be used in combination with optimized quarter sampling patterns such as those in \cite{Grosche2018}.
FSR has shown to be a successful reconstruction scheme for various inpainting and extrapolation tasks \cite{Herraiz2008, Stehle2006,Seiler2009} and showed best results for non-regular sampling and quarter sampling in~\cite{Schoberl2011,Seiler2015,Grosche2018}. 
An advancement of quarter sampling is called three-quarter sampling \cite{Seiler2018}. It increases the light sensitivity as well as the resolution per pixel. Here, more advanced reconstruction techniques such as the (local) joint-sparse  deconvolution and extrapolation (L-JSDE) need to be used \cite{Seiler2018, Grosche2020_localJSDE}. A three-quarter sampling sensor is illustrated in Figure\,\ref{fig:LR_vs_quarter}.

Another approach to acquire high resolution images without increasing the number of measured pixels is to upscale an image from a low-resolution sensor as it is also shown in Figure\,\ref{fig:LR_vs_quarter}. Unfortunately, using standard interpolation methods leads to a blurred and degraded image. A higher image quality can be achieved using super-resolution algorithms. During the last decade, great progress has been made in this field using sparse representation based approaches \cite{Yang2010} as well as convolutional neural network based approaches \cite{Dong2016}. 
Still one of the best approaches is the very deep super resolution (VDSR) network \cite{Kim2016}. It does not use a low-resolution image as input but enhances an image that was previously upscaled with bicubic interpolation (BIC) by learning the residuum between the unavailable high resolution image and the upscaled image. Using a rather simple deep network structure, VDSR achieves a high quality reconstruction especially at sharp edges.

Recently, a VDSR-like network has also been successfully used to increase the reconstruction quality for quarter sampling surpassing single-image super-resolution with VDSR. This approach is called VDSR-QS \cite{Grosche_VDSRQS}. It appends a VDSR-like network to an initial reconstruction with FSR. Additionally, a special data augmentation technique that is only possible for quarter sampling is used. As a downside, VDSR-QS requires more computation time than VDSR, because the initial reconstruction with FSR is computationally expensive and requires most of the processing time.

In this paper, we aim at increasing the reconstruction quality for (three-)quarter sampling even further. Therefore, we propose a novel end-to-end neural network that is capable of reconstructing different non-regular sensor layouts such as quarter sampling and three-quarter sampling.  Other than VDSR-QS, we do not require any classical reconstruction method as pre-processing making our approach purely neural network based.
 Effectively, we combine the theoretical advantages of quarter sampling compared to regular sampling with state of the art techniques from machine learning. Such a combination is different due to the non-regularity of the measurement process. Whereas a regular sampling, as it is done in single-image super-resolution, is a translationally invariant task, non-regular sampling is no longer translationally invariant. We compare our results with single image super-resolution as well as other (three-)quarter sampling reconstruction techniques.

This paper is organized as follows: In Section\,\ref{sec:vdsr_and_adapted_vdsr}, we briefly introduce the VDSR and VDSR-QS networks from literature. In Section\,\ref{sec:prop_method}, we present our newly designed neural network and explain how we deal with the non-existent translation invariance of the (three-)quarter sampling sensor measurements.
In Section\,\ref{sec:simulation_and_results}, we perform experiments that compare both processing chains and the used adaptations. We furthermore evaluate and discuss the results and provide visual examples.
In Section\,\ref{sec:conclusion}, we summarize the paper and give an outlook on future work.

\section{State-of-the-Art Implementations for Low-Resolution Sensors and Quarter Sampling Sensors}

\label{sec:vdsr_and_adapted_vdsr}

\subsection{VDSR for a Low Resolution Sensors}

\begin{figure}[t!]
	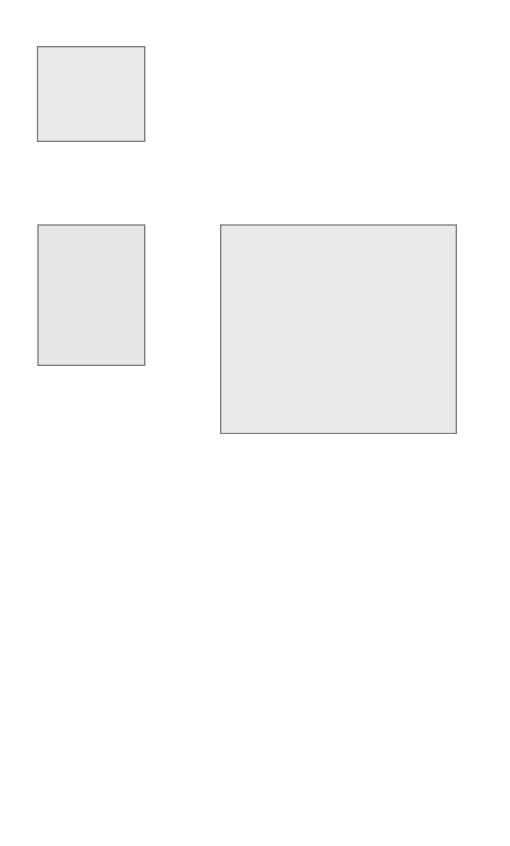
	\caption{Flow graph for sampling and reconstruction for (a) the low-resolution (LR) sensor with VDSR and (b) the quarter sampling (QS) sensor with VDSR-QS and (c) an arbitrary (non-regular) sampling sensor with the proposed locally fully connected reconstruction network.}
	\label{fig:reconstruction_network}
\end{figure}

For a low-resolution sensor as it is shown in Figure\,\ref{fig:LR_vs_quarter}, the acquisition of the image can be described as the filtering of a high resolution reference image $f_{ij}$ with a $2{\times}2$ filtering kernel of ones followed by a twofold sub-sampling in both spatial dimensions.
Effectively, each pixel measures the mean value of four high resolution pixels from a $2{\times}2$ neighborhood.
Prior to the application of a super-resolution algorithm such as the VDSR \cite{Kim2016}, the image is upscaled using bicubic interpolation leading to an approximate solution $\hat{f}_{ij}$.
In order to enhance the resolution of the upscaled image $\hat{f}_{ij}$, it is fed into VDSR \cite{Kim2016} being a convolution neural network trained to infer the residual $r_{ij} = f_{ij} - \hat{f}_{ij}$. The resulting image is calculated by summing the input and output of VDSR, i.e.,  
\begin{align*}
	\tilde{f}_{ij} = \hat{f}_{ij} + r_{ij}.
\end{align*}
The full processing chain for the case of the low-resolution sensor is shown in Figure\,\ref{fig:reconstruction_network}\,(a).

For the network architecture, we use a custom Tensorflow~\cite{tensorflow2015-whitepaper} and Keras~\cite{chollet2015keras} implementation of the original work in~\cite{Kim2016}. The same strategies and settings were used as in die original work expect for
the weights being adapted using the Adam optimizer~\cite{kingma2015} with an initial learning rate of $10^{-4}$ which is decreased by a factor of 10 after every 10-th epoch.
In total, the network is trained for 100 epochs. To increase the number of patches, and thus avoid overfitting during training, data augmentation is used by flipping and rotating the patches before feeding into the network.

\subsection{\mbox{VDSR-QS} for Quarter Sampling Sensors}

For the case of the quarter sampling sensor, a very similar processing chain is followed in \cite{Grosche_VDSRQS}. The non-regular sub-sampling of the quarter sampling sensor can be described as an element-wise multiplication of the reference image $f_{ij}$ with a binary mask $b_{ij}$. At this point, the sampled image $ (f_{ij} {\cdot} b_{ij})$ contains as many non-zero entries as pixels of the quarter sampling sensor. In order to reconstruct the missing pixels on the high-resolution grid, FSR \cite{Seiler2015} is used as initial reconstruction  $\hat{f}_{ij}$ . This reconstructed image is then fed into a VDSR network. Since FSR leaves the pixels that were actually measured with the quarter sampling sensor untouched, it is reasonable to set the residual at the output of the neural network to zero for those positions. The final image then reads 
\begin{align*}
	\tilde{f}_{ij} = \hat{f}_{ij} + r_{ij}\cdot(1-b_{ij}).
\end{align*}
This adaptation of VDSR is named \mbox{VDSR-QS} \cite{Grosche_VDSRQS} and is shown in Figure\,\ref{fig:reconstruction_network}\,(b).

For \mbox{VDSR-QS} the same set of hyper-parameters as in \cite{Grosche_VDSRQS} is used. Regarding the quarter sampling mask, we use the optimized quarter sampling mask from \cite{Grosche2018} because it shows an improved reconstruction quality for $\hat{f}_{ij}$. This mask is of size $32{\times}32$ pixels and is repeated periodically until the respective reference image is covered. Conveniently, such periodicity is beneficial for a future hardware implementation.

\section{Proposed Method}
\label{sec:prop_method}
In this paper, we propose to use an end-to-end neural network for non-regular sampling sensors. As input, the network uses the values measured on the sensor, e.g., a quarter sampling sensor, a three-quarter sampling sensor, or a low-resolution sensor. At the output, the network shall output the reconstructed high-resolution image. 
Our proposed network consists of two separate parts that are concatenated as shown in Figure\,\ref{fig:reconstruction_network}\,(c). The first part shall use the measurements from a fixed sensor and perform an initial reconstruction  $\hat{f}_{ij}$. This initial reconstruction is then fed into a standard VDSR network to further enhance the image quality.

The basic building block of neural networks such as VDSR are convolutional layers. These convolutional layers are translationally invariant in the sense that  shifted inputs lead to shifted outputs. Such translational invariance is not observed for a quarter sampling sensor due to the non-regularity of the sampling mask. In order to be able to meaningfully use convolutional neural networks, we need to re-introduce translational invariance into the sensor. This is achieved by using a sampling mask that repeats every $8{\times}8$ pixels. More precisely, we use an optimized mask of size $8{\times}8$ pixels from \cite{Grosche2018}. As a side effect, using periodically repeating masks can also be considered advantageous  for the hardware manufacturing.

As the mask repeats periodically, our task is reduced to the reconstruction of a $8{\times}8$ target block from the measurements. All other target blocks can then be reconstruction with the same operations. This is the case because the  relative geometric positions of the measurements are the same for each target block as a consequence of the periodicity of the mask.
Our neural network is designed to reconstruct all pixel values inside an $8{\times}8$ target block from the measurements geometrically lying inside this $8{\times}8$ block as well as those lying inside its surrounding support border with a width of 4 pixels. Effectively, this means that the measurements from within a $16{\times}16$ block are used to reconstruct all values inside its central $8{\times}8$ target block. For the next $8{\times}8$ target block, some of the measurements are used again since the support blocks overlap. This concept can be understood as a so-called sliding window reconstruction which recently proved to be useful even for arbitrary local measurements \cite{Grosche2020_localJSDE}.

The reconstruction of the $8^2=64$ pixel values in the target block shall be performed using several fully connected layers connecting them with the $16^2/4=64$ measurements in the support block. Fully connected layers are required due to the lack of translational invariance on this scale. Since the same fully connected layers are used locally for every target block, we call our approach a locally fully connected reconstruction (LFCR) network. \footnote{The source code of LFCR is published online:\\\texttt{https://gitlab.lms.tf.fau.de/LMS/lfcr\_public}}
In the implementation, we use a slightly smarter representation of the locally fully connected layers and avoid manual slicing and stacking such that a purely convolution neural network can be derived.

\begin{figure*}[t!]
	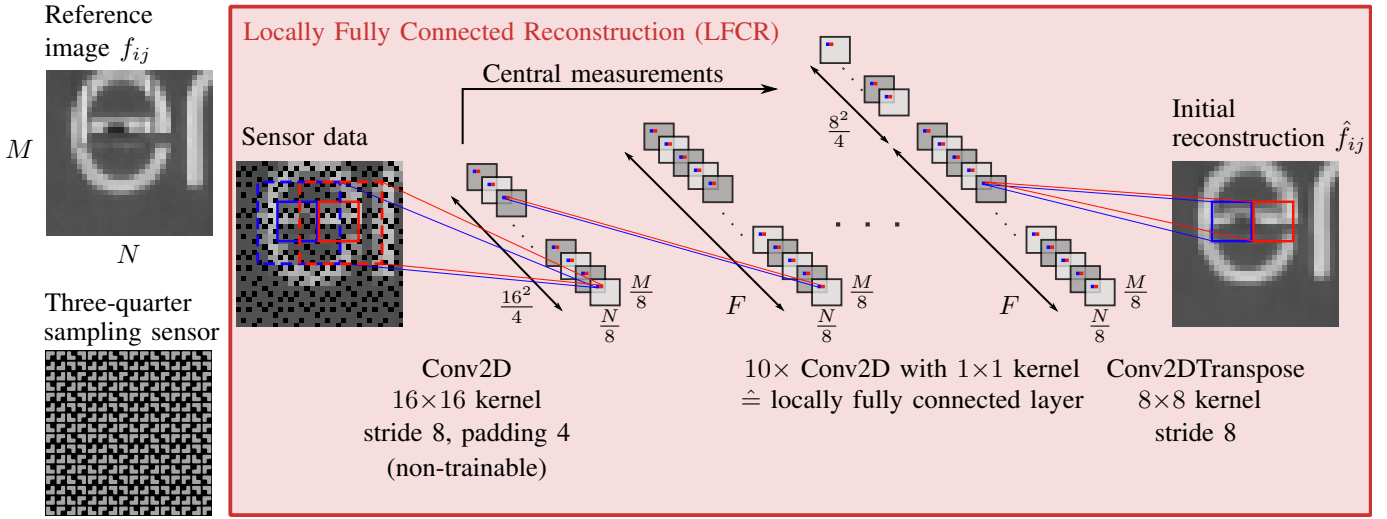
	\caption{Network layout of proposed LFCR network. First, a non-trainable convolutional layer is used to vectorize the measurements within a $16{\times}16$ support block as channels. The following convolutional layers with $1{\times}1$ kernel implement the locally fully connected link. Before the deconvolution, the central measurements are concatenated to the remaining channels. The bias as well as activation function are not illustrated for clarity.}
	\label{fig:network_first_half}
\end{figure*}

First, the measurements inside each $16{\times}16$ support block are accumulated in a vectorized form using a non-trainable convolutional layer with kernel size $16{\times}16$, depth $16^2/4=64$, stride $8{\times}8$, and padding 4. The convolution is applied to the reference image and the weights of the kernel are set such that a sampling with a \mbox{(three-)quarter} sampling sensor is mimicked. 
Next, the reconstruction of the $8^2=64$ values in the target block from the  $16^2/4=64$ measurements shall be performed using several fully connected layers. Each fully connected layer containing $F$ neurons can be written as a convolutional layer with kernel size of $1{\times}1$ and $F$ channels. The convolution is followed by a PReLu \cite{He2015_PReLu} activation function as non-linearity. The number of channels $F$, should be considerably larger than the number of missing pixels. We choose
\begin{align}
	F = 4 \cdot \underbrace{\frac{3}{4} 8^2}_{\substack{\text{Number of}\\{\text{missing pixels}}}} = 192,
\end{align}
where the additional factor of four gives the neural network enough freedom during training. After ten such convolutions, we concatenate the $8^2/4=16$ measurement channels from the beginning to the output. This way, the information such as the local mean value can be directly used by the de-convolution layer with stride  $8{\times}8$ and kernel size  $8{\times}8$ that is applied next. It uses the vectorized information at each position to find the pixel values in the target block of size $8{\times}8$ pixels. In the de-convolution, the target blocks are also  stacked next to each other forming an the output image $\hat{f}_{ij}$.
The complete structure of the proposed LFCR network is shown in Fig.\ref{fig:network_first_half}.

The just explained LFCR network is used as initial reconstruction network. Afterwards, its output is enhanced further using a VDSR-like network \cite{Kim2016}. The very deep convolutional layers  shall further improve the image quality. This idea is similar to  \cite{Kim2016, Grosche_VDSRQS}, where the outputs of classical upscaling/reconstruction algorithms are enhanced with VDSR. Slightly different to those works, we use the mean squared error as loss function, no kernel regularizer and the trainable PReLu activation function.

Overall, the LFCR+VDSR network features about $10^6$ trainable parameters. The first part (LFCR) uses $3.6\cdot 10^5$ parameters and the second part (VDSR) uses $6.6\cdot 10^5$ parameters. The entire network can be trained end-to-end. The two parts can be trained jointly as well as separately. For the loss function, we therefore define loss functions
\begin{align}
	&\mathcal{L}_\text{LFCR} =  \mathrm{MSE}(\hat{f}_{ij},f_{ij}), \\
	&\mathcal{L}_\text{LFCR+VDSR}=  \mathrm{MSE}(\tilde{f}_{ij},f_{ij})
\end{align}
for the intermediate reconstruction result after the first part and the final reconstruction result, respectively.
Training the two parts completely separately and sequentially was found to be more effective and resulted in a faster training and lower training losses. During the second training phase, the learning rate is reduced by a factor of ten compared to the initial learning rate since the network is already partially trained. An additional joint training phase reduced the training loss further. However, since the validation loss did not reduce further we omit the joint training in our simulations.

Similar to VDSR-QS \cite{Grosche_VDSRQS}, the non-regularity of the sensor layout allows for an additional dimension of data augmentation. Therefore, we  shift the reference image by 0, 2, 4, and 6 pixels in both spatial directions. While this results in different measurements for the non-regular sensor layouts, using such shifts for a low-resolution sensor is not meaningful. Effectively, shifting the reference images leads to an additional 16-fold  data augmentation for the given sensor layouts.

\section{Simulations and Results}

\label{sec:simulation_and_results}
In this section, we evaluate the performance of the different reconstruction networks for measurements using a quarter sampling sensor, a three-quarter sampling sensor, and a low-resolution sensor.

For training of the neural networks, we use custom TensorFlow \cite{tensorflow2015-whitepaper} implementations. Our implementations of VDSR and VDSR-QS were tested to show comparable results with respect to the original work in \cite{Kim2016, Grosche_VDSRQS}.

In terms of the training data, we use the same training data as in \cite{Kim2016, Grosche_VDSRQS}, i.e., the Set291 consisting of 291 images of various natural content. Slightly overlapping patches with a size of $41{\times}41$ pixels and  $48{\times}48$ were generated for the standard VDSR/VDSR-QS and our LFCR+VDSR network, respectively. The total number of measurements used for each case is comparable.
We simulate the monochrome sensors by converting all color images to grayscale with 8 bit depth scaled from $0\dots 255$.\footnote{Please note that some works in literature scale the data from $16\dots 235$ without taking such scaling into account during their PSNR calculations, systematically resulting in  1.32\,dB higher PSNR values.} Three-quarter sampling is simulated by averaging three high-resolution pixels as in \cite{Seiler2018}. All training patches are rotated and flipped such that an 8-fold data augmentation as achieved.

To evaluate the quality of the resulting images, we calculate the mean PSNR and the mean structural similarity (SSIM) \cite{Wang2004} for the images of the Urban 100 dataset from \cite{Huang2015} as well as for the Tecnick dataset \cite{Asuni2014}. Both datasets consist of natural image content. While the Tecnick images show a broad variety of content, the Urban100 images mainly focus on urban architecture.

In Table\,\ref{tab:results_first_half_vs_second_half}, we find that using the concatenation of the LFCR and the VDSR network results in more than 0.5\,dB higher PSNR values than only using the LFCR.
Secondly, Table\,\ref{tab:results_first_half_vs_second_half} confirms that using three-quarter sampling instead of quarter sampling results in higher reconstruction qualities not only with classical algorithms \cite{Seiler2018} but also with our neural network based reconstruction.
From now on, all results are therefore shown focusing on LFCR+VDSR reconstruction.
\begin{table}
	\caption{Comparison of using only the first part of our proposed network (LFCR) and both parts (LFCR+VDSR) for the reconstruction. The average PSNR in dB for the Urban 100 dataset is given for the two non-regular sampling scenarios, quarter sampling and three-quarter sampling.}
	\label{tab:results_first_half_vs_second_half}
	\centering
	\begin{tabular}{l|c|c}
		& Quarter  & Three-quarter \\
		& sampling &   sampling    \\\hline
		LFCR (only) &    27.61\,dB     &  28.45\,dB\\
		LFCR+VDSR &     28.13\,dB (+0.52\,dB)     & 29.12\,dB (+0.67\,dB) \\
	\end{tabular}
\end{table}

\begin{figure}[t]
	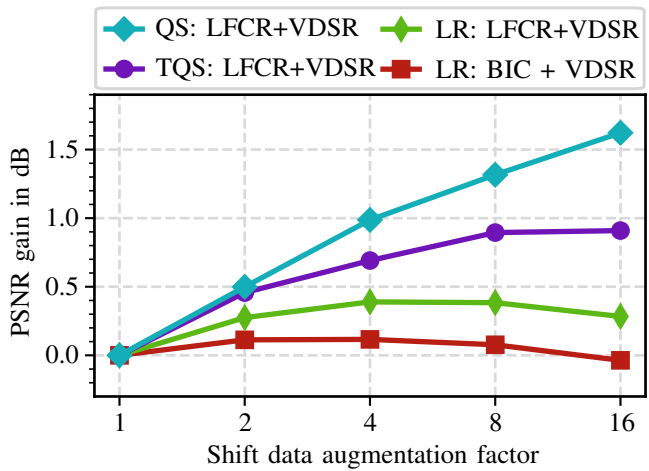
	\caption{Impact of different amounts of shift data augmentation. The PSNR gain in dB relative to the case without using the shift data augmentation is shown as a function of the shift data augmentation factor. The Urban 100 dataset is used. QS: Quarter sampling sensor, TQS: Three-quarter sampling sensor, LR: Low-resolution sensor.}
	\vspace*{-3mm}
	\label{fig:results_effect_shiftda}
	\vspace*{-2mm}
\end{figure}

Next, we analyze the impact of the special data augmentation technique that is possible for non-regular sampling cases as explained in Section\,\ref{sec:prop_method} and first proposed similarly in \cite{Grosche_VDSRQS}.
For this purpose, we performed additional trainings using reference images shifted by 0, 2, 4 and 6 pixels in both spatial dimensions. To understand the dependency on this kind of data augmentation, different amounts of such shifts are used and for each  amount we retrained our proposed network as well as the standard VDSR. As can be seen in Fig.\,\ref{fig:results_effect_shiftda}, using the shift data augmentation increases the reconstruction quality in terms of PSNR up to more than 1.5\,dB in case of quarter sampling and the Urban 100 dataset. Interestingly, for three-quarter sampling, a saturation is reached and using 8 or 16 different shifts only makes a small difference. For the low-resolution sensors, using the shifted inputs does not provide the networks with significant information. This explains the much smaller gains below 0.5\,dB for this scenario.
From now an, all results for the proposed LFCR+VDSR are shown for a shift data augmentation factor of 16, whereas the results for VDSR are shown without this data augmentation, because it is not used in \cite{Kim2016} and gains are marginal.

\begin{table}[t]
	\caption{Image quality in terms of average PSNR in dB and SSIM using the low-resolution sensor, the quarter sampling (QS) sensor and the three-quarter sampling sensor for the Urban100 and the Tecnick image dataset. Bold font indicates best PSNR/SSIM for each dataset.}
	\label{tab:results_VDSR_VDSRQS_ours}
	\centering
	\begin{tabular}{l|c|c}
		&            Urban 100             &             Tecnick              \\
		&           PSNR / SSIM            &           PSNR / SSIM            \\ \hline
		\hspace{-0.3em}Low-resolution sensor                  &                                  &                                  \\
		\,\,\,  BIC + VDSR \cite{Kim2016}                     &          28.92 / 0.9299          &          36.20 / 0.9746          \\
		\,\,\,  prop. LFCR (only)                             &          28.35 / 0.9243          &          35.86 / 0.9736          \\ 
		\,\,\,  prop. LFCR + VDSR                             &          28.73 / 0.9283          &          36.01 / 0.9739          \\ \hline
		\hspace{-0.3em}Quarter sampling sensor                &                                  &                                  \\
		\,\,\, FSR \cite{Seiler2015}, \cite{Grosche2018}      &          27.08 / 0.9116          &          34.11 / 0.9644          \\
		\,\,\, FSR + \mbox{VDSR-QS}   \cite{Grosche_VDSRQS}   &          29.29 / 0.9382          &          35.58 / 0.9709          \\
		\,\,\, prop. LFCR (only)                              &          28.65 / 0.9309          &          35.35 / 0.9698          \\
		\,\,\, prop. LFCR + VDSR                              &          29.76 / 0.9425          &          35.84 / 0.9720          \\ \hline
		\hspace{-0.3em}Three-quarter sampling sensor          &                                  &                                  \\
		\,\,\, L-JSDE \cite{Seiler2018,Grosche2020_localJSDE} &          27.09 / 0.9083          &         34.22 /  0.9654          \\
		\,\,\, prop. LFCR (only)                              &          29.47 / 0.9373          &          36.40 / 0.9751          \\
		\,\,\, prop. LFCR + VDSR                              & \textbf{30.03} / \textbf{0.9421} & \textbf{36.66} / \textbf{0.9758}
	\end{tabular}
	\vspace*{-3mm}
\end{table}
Using the insights from the above pre-evaluations, we compiled a comparison between the proposed LFCR and LFCR+VDSR (both with 16-fold shift data augmentation) and other (classical) reconstruction methods in Table\,\ref{tab:results_VDSR_VDSRQS_ours}. For this evaluation, we use the Tecnick dataset in addition to the Urban100 dataset.

From Table\,\ref{tab:results_VDSR_VDSRQS_ours}, we find that using three-quarter sampling in combination with the novel LFCR+VDSR neural network outperforms all other sensor layouts and reconstruction methods. Compared to a low-resolution sensor with VDSR, we observe a gain of 1.11\,dB and 0.46\,dB for the Urban100 and the Tecnick dataset, respectively.
Compared to the classical reconstruction algorithms (FSR and L-JSDE), we observe gains of 2.96\,dB and 2.44\,dB.
We would like to note that the key difference to single-image super-resolution is not the exact choice of the neural network but the underlying sampling method being non-regular.
The SSIM values also reported in Table\,\ref{tab:results_VDSR_VDSRQS_ours}  are in accordance with the PSNR results.

\begin{figure}[t!]
	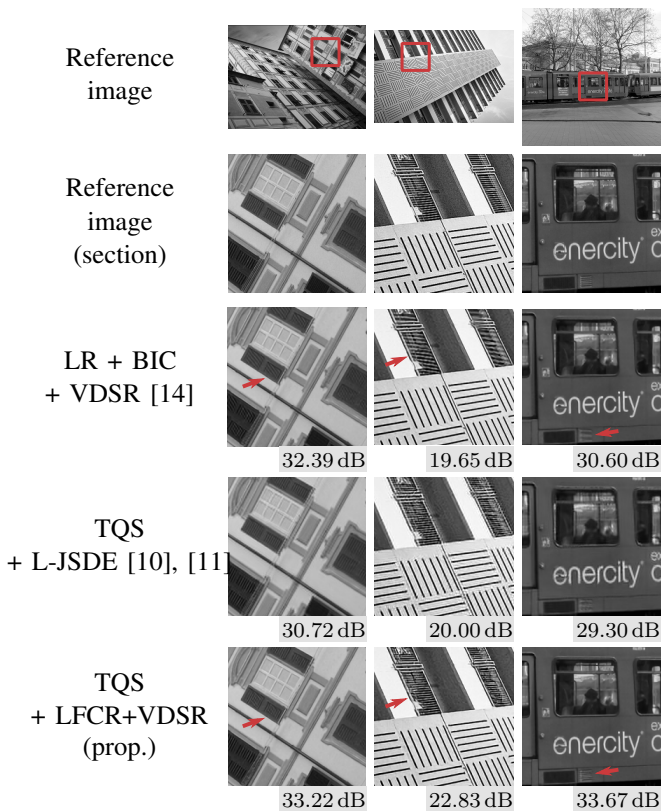
	\caption{Visual examples from sections of the Urban 100 dataset and the Tecnick. While the low-resolution (LR) sensor leads to severe aliasing, the three-quarter sampling sensor (TQS) enables higher visual quality in combination with the proposed LFCR+VDSR. \textit{(Please pay attention, additional aliasing may be caused by printing or scaling. Best to be viewed enlarged on a monitor.)}}
	\vspace*{-3mm}
	\label{fig:results_visual}
	\vspace*{-2mm}
\end{figure}
In Figure\,\ref{fig:results_visual}, we additionally show visual examples for two sections from the Urban 100 dataset and one section from the Tecnick dataset. Regions with high frequency content are affected from severe aliasing in case of the low-resolution sensor, regardless of using VDSR or not. Other than that, using three-quarter sampling in combination with the proposed LFCR+VDSR reduces the aliasing significantly. This can be seen especially for the window shutters (marker by a red arrow) as well as the ventilation slit of the train (marker by a red arrow).  We can also see that our neural network based approach outperforms the classical L-JSDE algorithm though L-JSDE can also make use of the non-regularity especially in case of very fine structures.

\vspace*{-1mm}
\section{Conclusion}
\label{sec:conclusion}
In this paper, we propose a novel end-to-end neural network approach consisting of two concatenated networks, the proposed LFCR and a standard VDSR. The combined LFCR+VDSR network is the first purely neural network based reconstruction approach for non-regular quarter sampling and three-quarter sampling sensor measurements. 
This approach outperforms other reconstruction techniques for the non-regular sampling methods and additionally outperforms data-driven approaches for single-image super-resolution. We combine the structural advantages of a non-regular sensor layout with the advantages of data-driven reconstruction. In designing the neural network, special care was taken to cope with the missing shift-invariance of the non-regular sampling sensors.

Using a three-quarter sampling sensor in combination with our proposed LFCR+VDSR network, we outperform the classical \mbox{L-JSDE} by 2.96\,dB on the Urban100 dataset. Compared to single-image super-resolution using a low-resolution sensor and VDSR, we gain 1.11\,dB for this dataset.
Future work may investigate the usage of other neural networks from single-image super-resolution in combination with LFCR and to expand LFCR to other (local) compressed sensing tasks, e.g. from \cite{Grosche2020_localJSDE}.

\vspace*{-0.3cm}
\bibliographystyle{IEEEbib}
\bibliography{literatur_jabref}

\begin{thebibliography}{10}

\bibitem{Schoberl2011}
Michael Sch{\"o}berl, J{\"u}rgen Seiler, Siegfried Foessel, and Andr{\'e} Kaup,
\newblock ``Increasing imaging resolution by covering your sensor,''
\newblock in {\em Proc. 18th {IEEE} International Conference on Image
  Processing}, Brussels, Sept. 2011, pp. 1897--1900.

\bibitem{Dippe1985}
Mark A.~Z. Dipp{\'{e}} and Erling~Henry Wold,
\newblock ``Antialiasing through stochastic sampling,''
\newblock in {\em Proc. 12th Annual Conference on Computer Graphics and
  Interactive Techniques}, New York, July 1985, pp. 69--78.

\bibitem{Hennenfent2007}
Gilles Hennenfent and Felix~J. Herrmann,
\newblock ``Irregular sampling: from aliasing to noise,''
\newblock in {\em Proc. 69th EAGE Conference and Exhibition}, London, June
  2007, pp. cp--27--00063.

\bibitem{Maeda2009}
Yui Maeda and Junichi Akita,
\newblock ``A {CMOS} image sensor with pseudorandom pixel placement for clear
  imaging,''
\newblock in {\em Proc. International Symposium on Intelligent Signal
  Processing and Communication Systems}, Kanazawa, Dec. 2009, pp. 367--370.

\bibitem{Seiler2015}
J{\"u}rgen Seiler, Markus Jonscher, Michael Sch{\"o}berl, and Andr{\'e} Kaup,
\newblock ``Resampling images to a regular grid from a non-regular subset of
  pixel positions using frequency selective reconstruction,''
\newblock {\em {IEEE} Transactions on Image Processing}, vol. 24, no. 11, pp.
  4540--4555, Nov. 2015.

\bibitem{Grosche2018}
Simon Grosche, J{\"u}rgen Seiler, and Andr{\'e} Kaup,
\newblock ``Iterative optimization of quarter sampling masks for non-regular
  sampling sensors,''
\newblock in {\em Proc. International Conference on Image Processing 2018},
  Athens, Oct. 2018, pp. 26--30.

\bibitem{Herraiz2008}
Joaquin~Lopez Herraiz, Samuel Espana, Esther Vicente, Elena Herranz, Manuel
  Desco, Juan~Jose Vaquero, and Jose Udias,
\newblock ``Frequency selective signal extrapolation for compensation of
  missing data in sinograms,''
\newblock in {\em Proc. {IEEE} Nuclear Science Symposium Conference Record},
  Dresden, Oct. 2008, pp. 4299--4302.

\bibitem{Stehle2006}
Thomas Stehle,
\newblock ``Removal of specular reflections in endoscopic images,''
\newblock {\em Acta Polytechnica}, vol. 46, no. 4, pp. 32, 2006.

\bibitem{Seiler2009}
J{\"u}rgen Seiler and Andr{\'e} Kaup,
\newblock ``Multiple selection extrapolation for improved spatial error
  concealment,''
\newblock in {\em Proc. {IEEE} International Workshop on Multimedia Signal
  Processing}, Rio de Janeiro, Oct. 2009, pp. 1--6.

\bibitem{Seiler2018}
J{\"u}rgen Seiler, Markus Jonscher, Thomas Ussmueller, and Andr{\'e} Kaup,
\newblock ``Increasing imaging resolution by non-regular sampling and joint
  sparse deconvolution and extrapolation,''
\newblock {\em {IEEE} Transactions on Circuits and Systems for Video
  Technology}, vol. 29, no. 2, pp. 308--322, Feb. 2019.

\bibitem{Grosche2020_localJSDE}
Simon Grosche, Andy Regensky, Jürgen Seiler, and André Kaup,
\newblock ``Boosting compressed sensing using local measurements and sliding
  window reconstruction,''
\newblock {\em {IEEE} Transactions on Image Processing}, vol. 29, pp.
  7931--7944, July 2020.

\bibitem{Yang2010}
Jianchao Yang, John Wright, Thomas~S Huang, and Yi~Ma,
\newblock ``Image super-resolution via sparse representation,''
\newblock {\em {IEEE} Transactions on Image Processing}, vol. 19, no. 11, pp.
  2861--2873, nov 2010.

\bibitem{Dong2016}
Chao Dong, Chen~Change Loy, Kaiming He, and Xiaoou Tang,
\newblock ``Image super-resolution using deep convolutional networks,''
\newblock {\em {IEEE} Transactions on Pattern Analysis and Machine
  Intelligence}, vol. 38, no. 2, pp. 295--307, feb 2016.

\bibitem{Kim2016}
Jiwon Kim, Jung~Kwon Lee, and Kyoung~Mu Lee,
\newblock ``Accurate image super-resolution using very deep convolutional
  networks,''
\newblock in {\em Proc. {IEEE} Conference on Computer Vision and Pattern
  Recognition}, Las Vegas, June 2016, pp. 1646--1654.

\bibitem{Grosche_VDSRQS}
Simon Grosche, Kristian Fischer, Fabian Brand, Jürgen Seiler, and Andre Kaup,
\newblock ``Enhanced image reconstruction from quarter sampling measurements
  using an adapted very deep super resolution network,''
\newblock in {\em Proc. IEEE International Conference on Image Processing}, Abu
  Dhabi, Oct. 2020, pp. 256--260.

\bibitem{tensorflow2015-whitepaper}
Mart\'{\i}n Abadi, Ashish Agarwal, Paul Barham, et~al.,
\newblock ``{TensorFlow}: Large-scale machine learning on heterogeneous
  systems,'' 2015,
\newblock Software available from \url{tensorflow.org}.

\bibitem{chollet2015keras}
Fran\c{c}ois Chollet et~al.,
\newblock ``Keras,'' \url{https://keras.io}, 2015.

\bibitem{kingma2015}
Diederik~P. Kingma and Jimmy Ba,
\newblock ``Adam: A method for stochastic optimization,''
\newblock in {\em Proc. International Conference on Learning Representations},
  San Diego, May 2015.

\bibitem{He2015_PReLu}
Kaiming He, Xiangyu Zhang, Shaoqing Ren, and Jian Sun,
\newblock ``Delving deep into rectifiers: Surpassing human-level performance on
  {ImageNet} classification,''
\newblock in {\em Proc. {IEEE} International Conference on Computer Vision},
  Dec. 2015.

\bibitem{Wang2004}
Zhou Wang, Alan~Conrad Bovik, Hamid~Rahim Sheikh, and Eero~P. Simoncelli,
\newblock ``Image quality assessment: From error visibility to structural
  similarity,''
\newblock {\em {IEEE} Transactions on Image Processing}, vol. 13, no. 4, pp.
  600--612, Apr. 2004.

\bibitem{Huang2015}
Jia-Bin Huang, Abhishek Singh, and Narendra Ahuja,
\newblock ``Single image super-resolution from transformed self-exemplars,''
\newblock in {\em Proc. {IEEE} Conference on Computer Vision and Pattern
  Recognition}, Boston, June 2015, pp. 5197--5206.

\bibitem{Asuni2014}
Nicola Asuni and Andrea Giachetti,
\newblock ``Testimages: a large-scale archive for testing visual devices and
  basic image processing algorithms,''
\newblock in {\em Proc. Smart Tools and Apps for Graphics - Eurographics
  Italian Chapter Conference}, Cagliari, Sept. 2014, pp. 63--70.

\end{thebibliography}

\end{document}